\documentclass{kluwer}   
\usepackage {graphicx}
\usepackage {longtable}

\begin{document}
\begin{article}

\begin{opening}
\title{\bf On solar neutrino fluxes in radiochemical experiments%
\footnote{The paper is extended version of the report presented 
on 9th International Pulkovo Conference on Solar Physics, 
Saint-Petersburg, Russia, July 4-9, 2005.}}

\author{R.N. Ikhsanov, Yu.N. Gnedin and E.V. Miletsky}
\runningauthor{R.N. Ikhsanov et al.} \runningtitle{On solar neutrino fluxes
} \institute{Central Astronomical Observatory of The
Russian Academy of Sciences at Pulkovo, 196140 St.Petersburg,
Russia (E-mail: solar1@gao.spb.ru)}
%\date{}

\begin{abstract}
We analyze fluctuations of the solar neutrino flux using data from
the Homestake, GALLEX, GNO, SAGE and Super Kamiokande experiments.
Spectral analysis and direct quantitative estimations show that
the most stable variation of the solar neutrino flux is a
quasi-five-year periodicity. The revised values of the mean solar
neutrino flux are presented in Table 4. They were used to estimate
the observed pp-flux of the solar electron neutrinos near the
Earth. We consider two alternative explanations for the origin of
a variable component of the solar neutrino deficit.

\end{abstract}

\end{opening}

\section{Introduction}

Investigation of solar neutrinos is one of the key directions of
modern astrophysics. The data from neutrino observatories (Super
Kamiokande, GALLEX,  SAGE and SNO) allowed to get important
information on both the solar neutrinos and the properties of
neutrino as an elementary particle. The observed solar neutrino
flux turned out to be by a factor of at least 2 smaller than
predictions of the standard solar model (SSM). The currently most
commonly accepted mechanism responsible for the solar neutrino
deficit is the process of neutrino oscillations whereby electron
neutrinos may be converted into either muon or tau neutrinos. This
transformation is the most likely to be resulted from the effect
of Mikheev, Smirnov and Wolfenstein (MSW) \cite{Mikheev} 
taking place as neutrinos propagate through matter in the
solar interior. The MSW effect is
based on the differences in interaction with matter for neutrinos
of different types.

Solar neutrino studies have been repeatedly reviewed. The latest
detailed review of the solar neutrino problem was presented by
Bahcall \cite{Bahcall04}. New important results have been obtained in
the recent experiments with anti-neutrino from the ground-based
reactors (the KamLand experiment \cite{KamLAND}).

The total deficit of the solar neutrino flux as compared to SSM
can be presented as a sum of two components: a constant component,
caused purely by neutrino oscillations, and a variable one,
connected presumably with solar activity and, in particular, with
magnetic field variations in the interior of the Sun.

A physical mechanism, responsible for such kind of variability can
be a so called ``resonant spin-flavor precession'' (RSFP, see,
e.g., \cite{Schechter, Akhmedov, Lim}), which involves a nonzero 
magnetic moment of neutrino. In
this case, a strong magnetic field, which probably exists at the
base of the solar convective zone, causes a strong spin
precession, which converts some of the left-handed electron
neutrinos, produced by nuclear reactions in the core, into sterile
right-hand neutrinos which do not participate in the weak
interaction. The stronger magnetic field is, the more probable is
neutrino transition from one state to another, resulting in
reducing the number of neutrinos of certain type at the surface of
the Earth. Although the magnetic moment of a Dirac neutrino is
essentially small ($\mu_\nu \approx 3.1\cdot 10^{-19} \mu_{\rm B}$) 
due to quantum radiative corrections, there
exist models, in which the value of neutrino magnetic moment is in
the interval from $10^{-10}$ to $10^{-11}\,\mu_{\rm B}$ 
\cite{Klapdor}.

Due to admixture of the strong and electromagnetic interactions to
the week interaction in the super-symmetry theory there is no
direct dependence between the magnetic moment and the mass of
neutrino. An interesting variant arises in the SUSY theory with
SU(2) symmetry, in which the right- and left-hand neutrinos form a
close doublet. In this situation the neutrino magnetic moment is
non-zero even in the case of zero mass. Theoretical studies have
shown that if the mass of electron and muon neutrino is under
10\,eV, the magnetic moment of neutrino of both types is $\sim
10^{-11}$---$10^{-10}\,\mu_{\rm B}$. This value of magnetic moment
can affect the solar neutrino experiments, especially taking into
account that the magnetic field in the Sun's interior can exceed
$10^7$\,G at the distance of 0.2 solar radius 
\cite{Couvidat, Balantekin}.

Although an existence of a variable component of the solar
neutrino flux and its possible connection with solar activity
remains under discussion for many years, the ultimate conclusion
has not been made so far. A number of authors have reported this
connection  (see, e.g., reviews \cite{Ikhsanov02, Ikhsanov03}),
 while the other criticize such claims --- specifically the
claim that the neutrino flux is anti-correlated with some indices
of solar activity \cite{Klapdor}.

A major part of recent studies have not revealed variations of the
solar neutrino flux \cite{Cattaneo, Yoo, Pandola}, but some authors 
still report observations of the variable component of the solar
neutrino flux \cite{Milsztajn, Caldwell}. The latter concluded about possible
contribution of the spin-flavor process to the variable part of
the solar neutrino deficit.

Searching for associations with the solar activity cycle, most
investigations focused on correlation (anti-correlation) with the
11- and 2-year periodicity. Ikhsanov and Miletsky 
\cite{Ikhsanov99, Ikhsanov02}
have shown that  a periodicity close to 5 years  plays the most
important role in the variations of the neutrino flux. They have
made an attempt to find indices of solar activity, variations of
which show pronounced manifestation of the quasi-5-year
periodicity.

 In this paper we address some pros and cons of the neutrino flux
 deficit variations with the periods in excess of one year.
First we examine the data from neutrino experiments by means of
spectrum analysis and show that the data from Homestake and GALLEX
exhibit flux fluctuations at the $2\sigma$-level, while in the
data from GNO and SAGE these oscillations are within errors. Then
we make quantitative estimates of the neutrino flux variations and
discuss their possible relation to the solar activity. We argue
that the mean flux estimates from the radiochemical experiments
need correction and present the revised values in Table\,4. They
were used to evaluate the observed pp-flux of the electron
neutrino at the surface of the Earth.

\section{Variations of the solar neutrino flux}

We use the data, obtained with the Cl-Ar detector Homestake
\cite{Cleveland} with a threshold value of 0.814 MeV,  the
data from the observatories GALLEX, GNO \cite{Kirsten, Pandola}
and SAGE \cite{Abdurashitov} working with 
gallium-germanium detector and a threshold
value of 0.233 MeV, as well as Super-Kamiokande data \cite{Yoo}, with
a threshold value of 7 MeV and averaged for 10 days.

Fig.\,1a shows the power density spectra (PDS) for the neutrino
flux (or, more exactly, for the $^{37}$Ar counting rates) from the
Homestake experiment, calculated using the Lomb-Scargle technique
\cite{Scargle}. Two peaks with the periods of 4.6 and 2.1 years can be
distinguished at the level of confidence probability equal to or a
bit higher than 95\% 
(Table\,1). The levels of confidence
probability of the most pronounced peaks were determined by means
of shuffle technique \cite{Sturrock}, with 10\,000 test
spectra having been computed for each estimate. This evaluation of
confidence allows to take off some limitations of standard methods
being in use in this kind of investigations. For example, a
criterion, which was  applied in some recent papers \cite{Yoo, Pandola},
significantly diminish the confidence level of the revealed
periods. It should be noted here that statistical criteria being
used in these studies contain assumptions about stationarity and
normality of the data, while the processes under investigation are
non-stationery and quasi-periodical, and, hence, require more
refine and adequate criteria of significance, especially if the
amplitudes of quasi-periodical variations are relatively small. In
such cases it makes sense to perform additional analysis by means
of heuristic methods.

\begin{figure}
\begin{center}
%\framebox{
\includegraphics[width=0.95\textwidth,trim=15 10 55 10,clip]{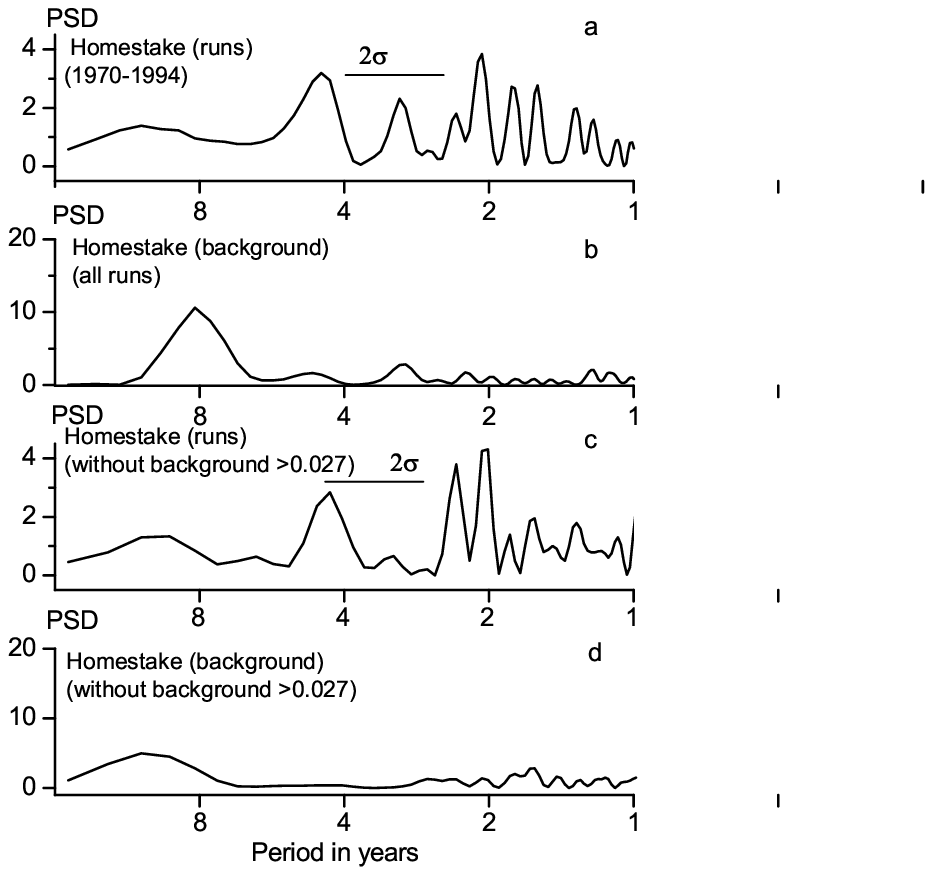} %0.84
%}
\caption{
The Power Spectral Density (PSD) for the neutrino flux from the Homestake 
experiment: (a) - for all runs; (b) - for background values; 
(c) -  for all runs except those which have background values 
more then 0.027 at/day; (d) - for background values with 
values less then 0.027 at/day.}
\label{fig1}
\end{center}
\end{figure}

\begin{table}
%\begin{center}
\begin{tabular}{|c|c|c|c|}
\hline
Period (years)& 
10.6& 
4.6& 
2.1 \\
\hline
Power Spectral Density& 
1.4& 
3.2& 
3.8 \\
\hline
Level of confidence probability ({\%})& 
67& 
96& 
97 \\
\hline
\end{tabular}
\caption
{
Confidence probability levels of greatest maximums on the neutrino 
flux Power Spectral Density (PSD) from the Homestake experiment.
}
\end{table}

Let us note that numerous peaks around a 2-year period are caused
by its instability \cite{Ikhsanov02, Ikhsanov03}, and a
duration of solar neutrino data is not long enough to permit a
reliable detection of 11-year periodicity. In order to check, to
what extent the character of neutrino fluctuations  is affected by
the background value, we have removed the runs with a large
background level ($>0.027$\,count per day) from the Homestake
series (about 1/3 of total number) \cite{Cattaneo}. The power spectrum was
then calculated. The result, shown in
Fig.\,1\,b,c, allows to conclude that the influence of the
background value is not essential: the same peaks, corresponding
to 10, 4.6 and 2 years are seen in the PSD.

\begin{table}
%\begin{center}
%\begin{tabular}{|c|c|c|p{49pt}|p{88pt}|p{49pt}|}
\begin{tabular}{|c|c|c|c|c|c|}
\hline
\textbf{No}& 
\textbf{Date}& 
\textbf{Maximum}& 
\textbf{Number}& 
\textbf{Minimum}& 
\textbf{Number}\\
\textbf{}& 
\textbf{}& 
\textbf{at/day}& 
\textbf{of runs} \par \textbf{(n)}& 
\textbf{at/day}& 
\textbf{of runs} \par \textbf{(n)} \\
\hline
1a& 
1970.5& 
-& 
& 
0.27 $\pm $ 0.20(?)& 
4 \\
\hline
1b& 
1970.5& 
-& 
& 
0.21 $\pm $ 0.19 (?)& 
3 \\
\hline
2a& 
1973.0& 
0.49 $\pm $ 0.47(?)& 
5& 
-& 
 \\
\hline
2b& 
1973.0& 
0.62 $\pm $ 0.59(?)& 
3& 
-& 
 \\
\hline
3a& 
1975.0& 
-& 
& 
0.44 $\pm $ 0.31& 
6 \\
\hline
3b& 
1975.0& 
-& 
& 
0.38 $\pm $ 0.15& 
3 \\
\hline
4a& 
1977.8& 
0.72 $\pm $ 0.34& 
8& 
-& 
 \\
\hline
4b& 
1977.8& 
1.00 $\pm $ 0.13& 
3& 
-& 
 \\
\hline
5a& 
1980.2& 
-& 
& 
0.17 $\pm $ 0.19(?)& 
8 \\
\hline
5b& 
1980.2& 
-& 
& 
0.02 $\pm $ 0.03(?)& 
3 \\
\hline
6a& 
1982.3& 
0.54 $\pm $ 0.34& 
8& 
-& 
 \\
\hline
6b& 
1982.3& 
0.65 $\pm $ 0.14& 
3& 
-& 
 \\
\hline
7a& 
1984.3& 
-& 
& 
0.48 $\pm $ 0.34& 
9 \\
\hline
7b& 
1984.3& 
-& 
& 
0.31 $\pm $ 0.11& 
3 \\
\hline
8a& 
1986.8& 
0.80 $\pm $ 0.32& 
4& 
-& 
 \\
\hline
8b& 
1986.8& 
0.85 $\pm $ 0.18& 
3& 
-& 
 \\
\hline
9a& 
1989.0& 
-& 
& 
0.49 $\pm $ 0.30& 
9 \\
\hline
9b& 
1989.0& 
-& 
& 
0.28 $\pm $ 0.24& 
3 \\
\hline
10a& 
1991.5& 
0.59 $\pm $ 0.24& 
8& 
-& 
 \\
\hline
10b& 
1991.5& 
0.75 $\pm $ 0.20& 
3& 
-& 
 \\
\hline
Mean& 
a& 
0.66 $\pm $ 0.31& 
28& 
0.47 $\pm $ 0.32& 
24 \\
\hline
Mean& 
b& 
0.77 $\pm $ 0.17& 
12& 
0.32 $\pm $ 0.18& 
9 \\
\hline
Mean& 
a& 
\multicolumn{3}{|c|}{0.566 $\pm $ 0.313} & 
52 \\
\hline
Mean& 
b& 
\multicolumn{3}{|c|}{0.545 $\pm $ 0.173} & 
21 \\
\hline
\end{tabular}
\caption{
The  maximum and minimum values of the counting rates 
from the Homestake experiment.  
}
%\end{center}
\end{table}

A short duration of data series from the GALEX+GNO, SAGE
and SK experiments does not allow to reveal periodicities longer
than 2 years in their PSD. Fig.\,2a shows that PSD of a whole GAL+GNO
data series exhibits only three modes in excess of $2\sigma$:
1.25, 2.34 and 4.1 years. Separate presentation of the PSD for the
series GAL (Fig.\,2b) and GNO (Fig.\,2c) indicates that these
peaks are seen only in the first PSD, with the 2.2-year period
being revealed with a large confidence ($\sim 99.5\%$). It should
be noted that three peaks of quasi-two-year oscillations are
clearly seen on the graph representing the original data series
(see Fig.\,1 \cite{Pandola}). The PSD of the GNO series does not show
any significant periods, while the PSD of the SK series (for the
same time interval) exhibits  two peaks at a $2\sigma$-level: 2.6
and 1.22 years.

\begin{figure}
\begin{center}
%\framebox{
\includegraphics[width=0.65\textwidth,trim=15 10 35 10,clip]{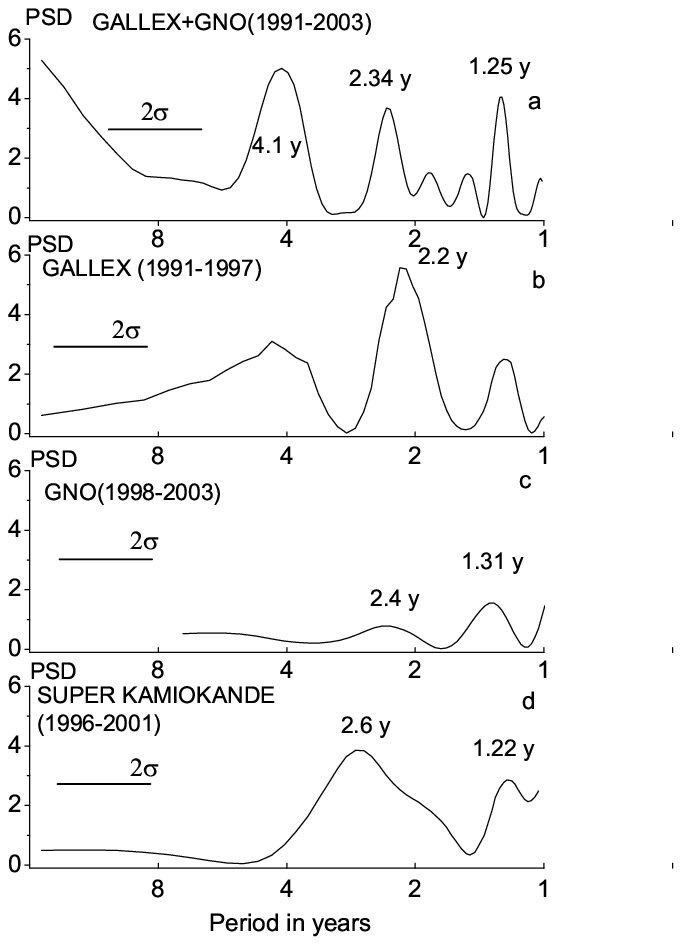} %0.84
%}
\caption{
The Power Spectral Density (PSD) for the neutrino flux from experiment:
(a), (b), (c) -  GALLEX and GNO;  (d) - Super Camiokande.
}\label{fig2}
\end{center}
\end{figure}

Another picture is seen in the SAGE experiment. In the PSD for a
whole data series (Fig.\,3a) only one peak at 2.1 year can be
suspected, and it is noticeably below a $2\sigma$-level. Separate
presentations of the PSD for the time intervals 1990-1996 and
1997-2001 show no significant peaks either, although the
quasi-2-year peaks are in the latter case even higher than in the
Homestake data for 1989-1994 (Fig.\,3d). The above analysis allows
to conclude that detection of a 2-year quasi-periodicity in the
neutrino flux data is unstable and hardly reliable.

\begin{figure}
\begin{center}
%\framebox{
\includegraphics[width=0.65\textwidth,trim=15 10 35 10,clip]{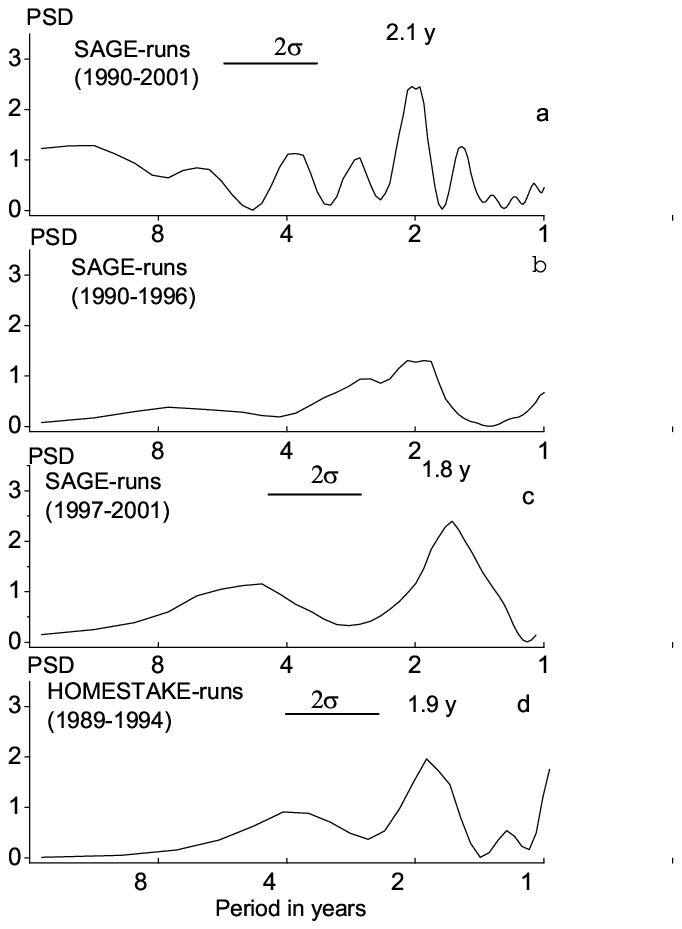} %0.84
%\includegraphics[width=0.95\textwidth,trim=15 15 30 20,clip]{fig3s.eps} %0.84
%}
\caption{
The Power Spectral Density (PSD) for the neutrino flux from experiment:
(a), (b), (c) -  SAGE ;  (d) - Homestake.
}\label{fig3}
\end{center}
\end{figure}

Thus, the data from the Homestake experiment indicate that a
quasi-5-year period is the most stable in the solar neutrino flux
fluctuations. Let us consider pros and cons of reliability of this
periodicity.

1. The smoothing of the Homestake series using the broadband
Butterworth filter (with the frequency interval of
maximum transmission, corresponding to the period interval 4-14
years) \cite{Ikhsanov99} reveals six consequent 
peaks on a 5-year wave (with account
for GALLEX data) with the average distance between maxima of about
$4.6\pm 0.7$ years (Fig. 4b). As can be seen from this figure, the
distance between the centers of neighboring hills changes from 3.9
to 5.3 years, i.e. the neutrino flux variations do show a 5-year
periodicity.

2. Table\,2 presents the average values of the counting rates from
the Homestake experiment for all the runs within $\pm 0.75$\,years
from the centers of maxima and minima of the flux fluctuations
from Fig.\,4b (case ``a''). Besides of that, the same calculations
have been done for the case ``b'', when only three consecutive
runs have been taken within the same limits (with the maximum and
minimum flux values respectively). The statistical errors on the
confidence level of 68\% ($1\sigma$) are also given here. The
values marked by ``?'' have statistical errors at the level of
flux estimate itself. They were not taken into account when
calculating the column mean values. The following conclusions can
be made examining the Table\,2:

Firstly, all flux values from the column of maxima are higher than
minimum values next to them. In the case (b), the difference
between neighboring minima and maxima are mostly exceeds of
$2\sigma$, and is on average $2.6\sigma$, that coincides with the
confidence level, determined from the power spectrum (Table\,1).
In the case (a) this difference is significantly smaller -
$0.6\sigma$ on average.

Secondly, the maximum values of the neutrino flux can be divided
into two groups: high values N\,4,8,12 and low values N\,2,6,10.
The column mean value of the maximum neutrino flux is $0.66\pm
0.31$ at/day, that corresponds to 46\% of the SSM value (1.44
at/day), while the maxima N4,8 constitute 53\% of the SSM
value. Taking into account that according to Table\,2 the mean
minimum value of the neutrino flux is $0.47\pm 0.32$ at/day,
that is 33\% 
of the SSM value, the modulation of the flux
fluctuations can be as high as 20\% 
of the SSM value. However,
this value could be considered as statistically justified only if
statistical errors given above were at least by a factor of 3
smaller than those from the Homestake experiment. That is why the
only statistically significant estimate from the Homestake
experiment is the mean value of neutrino flux, determined as an
average of the mean maximum and minimum values from Table\,2. It
equals $0.565\pm 0.044$ at/day, that is $0.392\pm 0.030$ of the
SSM value. The error was estimated as a root-mean-square deviation
from the mean ($\sigma/\sqrt{n}$)
for $n=52$. We could leave off at this point, as
many other authors considering this question. But the fact that a
five-year fluctuation of the neutrino flux manifests itself
continuously throughout the whole period of observations
(Fig.\,4b) allows to claim the reality of this quasi-periodicity
in the Homestake series.

Thirdly, a comparison of the maxima positions on the neutrino flux
curve (Fig.\,4b) with the cyclic variations of the large-scale
solar magnetic field strength on the transit of neutrinos from the
solar core to the Earth (Fig.\,4a) reveals a certain regularity.
The positions of high maxima (N 4,8,12) coincide with the phases
of minima of the 11-year cycle in the equatorial zone of the
large-scale magnetic field. The maxima of smaller height (N 6,10)
were observed during the change of polarity of the large-scale
magnetic field in the equatorial zone, while the minima of the
neutrino flux (N 5,9) occurred during the periods of maxima of the
11-year cycle. Thus, the Homestake series shows fluctuations with
the quasi-periods of about 10 and 5 years, thus demonstrating
their interconnection with the cyclic changes of the solar
magnetic field. Such regularity can hardly be an accidental
phenomenon.

\begin{figure}
\begin{center}
%\framebox{
\includegraphics[width=0.75\textwidth,trim=15 10 35 10,clip]{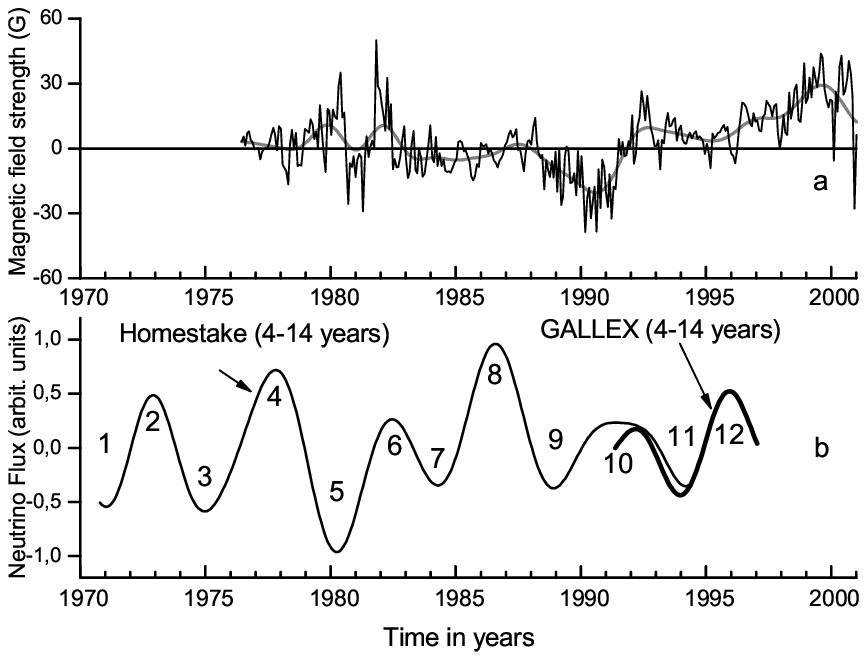} %0.84
%\includegraphics[width=0.95\textwidth,trim=15 10 55 10,clip]{fig4s.eps} %0.84
%}
\caption{Time variations of the large-scale solar magnetic field strength on the transit of neutrinos from the
solar core to the Earth - (a); Time variations neutrino flux from Homestake and GALLEX 
experiments, after procedure of Butterworth filtration with the frequency interval of maximum 
transmission, corresponding to the period interval 4-14.
}\label{fig4}
\end{center}
\end{figure}

This problem could be solved by comparison with the data from
other independent experiments. Unfortunately, observations with
the Homestake experiment finished at the first set-out of other
neutrino experiments. However, the data from GAL+GNO can be
considered as some resumption of the Homestake observations: they
show Homestake-like PSD (Fig. 1a, 2a) and a presence of
quasi-5-year hills. 

\begin{table}
\begin{tabular}{|c|c|p{57pt}|p{28pt}|p{57pt}|p{28pt}|p{42pt}|}
\hline
No& 
Date& 
Maximum \par (data in \par SSM units)& 
Number \par of runs \par (n)& 
Minimum \par (data in \par SSM units)& 
Number \par of runs \par (n)& 
Experiment  \\
\hline
1& 
1991.8& 
0.82 $\pm $ 0.26 \par 0.59 $\pm $ 0.62(?)& 
9 \par 9& 
& 
& 
GALLEX \par SAGE \\
\hline
2& 
1994.2& 
& 
& 
0.59 $\pm $ 0.27 \par 0.79 $\pm $ 0.54(?)& 
15 \par 10& 
GALLEX \par SAGE \\
\hline
3& 
1996.6& 
0.99 $\pm $ 0.28 \par 0.42 $\pm $ 0.15 \par 0.44 $\pm $ 0.05& 
11 \par 7 \par 22& 
& 
& 
GALLEX \par SAGE \par SK \\
        \hline
4& 
2000.2& 
& 
& 
0.41 $\pm $ 0.06 \par 0.58 $\pm $ 0.34 \par 0.46 $\pm $ 0.05& 
7 \par 9 \par 17& 
GNO \par SAGE \par SK \\
\hline
5& 
2001.8& 
0.61 $\pm $ 0.24 \par & 
7 \par & 
& 
& 
GNO \par  \\
\hline
\end{tabular}
\caption{
The mean neutrino flux values from the GALLEX, GNO, SAGE 
and Super Camiokande (SK) experiments.
}
%\end{center}
\end{table}

Similar to Table\,2, Table\,3 presents the
neutrino flux values GAL+GNO and their standard deviations
within the time intervals, which have been chosen shorter than in
Table\,2 (0.6--0.9 year) in order to get approximately the same
number of runs. The centers of these time intervals are given in
the second column of Table\,3. The flux values from the SAGE and
SK experiments during the same time intervals are presented in the
second and third rows, respectively. For the sake of further
comparison, the fluxes have been normalized to the corresponding
SSM-values. Similarly to the Homestake data, Table\,3 shows that
the GAL+GNO count rates (flux values) from the column ``Maximum''
are in all the cases higher than the neighboring minima. This
difference is particularly noticeable between the fluxes at the
phases of minimum (N3) and maximum (N4) of the solar cycle -
$3.4\sigma$. However, the data from SAGE experiment shows an
opposite behavior during the same time intervals - the values from
the column ``Minimum'' are higher than the values from the column
``Maximum'', although the difference between them remains within
errors. What is the reason for so significant divergency of
results obtained from the same Gallium experiment? First of all,
it is connected with large observational errors. Table\,3 shows
that the SAGE flux estimates before 1996 (hereafter, SAGE I) are
contaminated with errors at the level of or even higher than the
flux values themselves (N1). Such cases are marked by ``?'' (as in
Table\,2). However, from 1996 onwards (Sage II) the errors are at
the level compared to that of the GAL+GNO and Homestake experiments.

Thus, Table\,3 reveals that a comparison of the neutrino flux at
the phases of minimum and maximum (N4 and N3, respectively) allows
to conclude that the series SAGE and GAL+GNO show connection with
solar activity, although within errors.

\section{Discussion}

Studies of different periodicities in the solar activity (from the
index of sunspots areas and Wolf numbers to the global magnetic
field of the Sun) in the range 1--11 years 
(see \cite{Ikhsanov96, Ikhsanov98} and references therein) 
have shown that in addition to
the 11-year cycle there exist three quasi-periodic modes centered
at 1.3, 2.1 and 4.4 years. As can be concluded from the Figures
1--4 and Table\,2, these and only these quasi-periodic modes have
been observed in the variations of the solar neutrino flux. A
natural question arises, if the observed fluctuations result from
a real change of the neutrino flux or they are caused by an
exterior effect on the experiment, operating synchronously with
solar activity.

If the Homestake data really reflect the value of incoming
neutrino flux from the Sun, then, according to Table\,2, a total
deficit of the neutrino flux consists of constant and variable
components. The constant component of the flux deficit can be
determined as a difference between the estimated SSM flux and the
maximal observed flux, that is about 50\%, according to Table\,2.
Then a variable component can constitute about 20\% of the SSM
value. According to recent theoretical and experimental studies,
the constant component of the flux deficit can be explained in the
frame of neutrino oscillations theory (MSW-effect) 
\cite{Mikheev, Klapdor},
while interpretation of the second (minor) component of the flux
deficit requires implementation of the mechanism accounting for
variations of the solar magnetic field. Such mechanism is likely
to be a ``resonant spin-flavor precession'' (RSFP), which implies
a significant magnetic moment of neutrino.

However, two parallel observational series of the Ga-experiment,
SAGE and GAL+GNO (see Table\,3), provide controversial results on
the variable component of the neutrino flux deficit. Let us note,
that in spite of relative scarceness of the data (1990-2003),
Table\,3 allows to make some important conclusions. As mentioned
above, the table reveals that during the time interval 1990-1996
(N1,2) the errors in the SAGE data are very significant. The same
can be seen from the mean flux value calculated within this period
of time : $87.8\pm12.4$\,SNU. However, this flux value and large
errors are caused presumably by the runs, in which the flux
estimate exceeded the mean value by a factor of 2-3. After removal
of these known too high estimates of 5 runs, one finds SAGE\,I to
be $67.1\pm 7.9$\,SNU ($N=39$), or $0.520\pm 0.061$\,SSM. Although
a root-mean-square error remains large, the mean flux value is in
good agreement with the mean flux of SAGE\,II - $66.1\pm 5.3$\,SNU
($0.512\pm 0.041$\,SSM), calculated for the time interval
1996--2001 (48 runs). Taking into account that in 1996 the
observing programm was changed \cite{Abdurashitov}, 
SAGE\,I and SAGE\,II can be
considered as two independent series.

Addressing Table\,3 again, we find that an averaged counting rate,
determined as a half-sum of the column-mean values for ``Maximum''
and ``Minimum'', is about 0.50\,SSM for SAGE\,II and GNO, while
for the GALLEX series it is 0.75\,SSM, i.e. the latter case shows
a significant shift of the zero point with respect to the SAGE and
GNO data. According to the SAGE data, the constant component of
the flux deficit is of the order of 50\%, while a variable
component, in contrast to GALLEX, is within errors. In this
connection, it seems necessary to consider an alternative reason
for the variable deficit component to arise in the Homestake and
GALLEX data, namely, an unproper background reduction.

As shown by Cattaneo \cite{Cattaneo}, the neutrino flux values from the Homestake
data are strictly correlated with the background level. His main
conclusion was that the runs with high background level are
unreliable and should be excluded from the further analysis.
Having removed such runs (about one third of the total amount), he
managed to increase a mean neutrino flux in the Homestake
experiment up to $0.566\pm 0.030$ atom/day. The removed 30\% of
the runs belong predominantly to the first half of the series,
while the second half displays the neutrino flux which is on
average systematically higher by approximately the same value. It
is clearly seen on the low flux values and large errors in the
case N1, and particularly N5 of Table\,2, and just in these cases
the background magnitude is in excess of 0.027 at/day. The
derived above mean counting rate of the solar neutrino in the
Cl-experiment ($0.565\pm 0.044$\,atom per day, according to
Table\,2) is in good agreement with the value obtained by
Cattaneo, while the commonly adopted value $0.478\pm
0.03$\,atom/day, presented by Clevelend \cite{Cleveland}, is
obviously underestimated and corresponds to the minimal flux
derived from Table\,2.

However, as seen in Fig.\,1c and Table\,2, even a new mean flux
estimate does not essentially change a shape of PSD, that is it
still shows neutrino flux fluctuations, corresponding to
quasi-periods of 10, 4.6 and 2 years. That is why it seems
reasonable to unite a statement about a background level with that
about its dependence on the solar activity.

Really, analogous to the Homestake experiment, the variable flux
component in the SAGE and GNO observations contributes about 0.20
of the SSM value, but in the GAL+GNO experiment the maximum neutrino
flux was observed at the phase of minimum of the 11-year solar
activity cycle, while in the SAGE experiment - at the phase of
maximum. This means that flux fluctuations are likely to be
connected with an opposite reaction of these two experiments on
the influence of solar activity, rather than with a real change of
the solar neutrino flux. Then the amplitude of these fluctuations
of the neutrino flux, depending on the phase of the 11-year solar
activity cycle, lies within the limits $\pm 0.1$\,SSM relative to
the mean solar neutrino flux. In this case the maximum neutrino
flux in the Cl- and Ga-experiments is likely to be about 3.83\,SNU
and 77.6\,SNU, respectively. The PSD on Fig.\,2d evidences that an
influence of this unknown factor, that is an influence of solar
activity on the experimental results, may extend over the
SK-experiment as well.

According to Vladimirsky and Bruns \cite{Vladimirsky}, this factor can be
connected with the influence of geophysical factors, controlled by
the solar activity, on the physicochemical kinetics of the target
material.

The GALLEX data from Table\,3 display the same picture as the
Homestake data, with the flux value from the column ``Minimum'',
$0.59\pm 0.27$\,SSM,  being significantly higher than the mean
flux estimate in the SAGE and GNO experiments. This means that the
GALLEX series should not be used in calculation of the mean value
over all the series of the Ga-experiment.

Table\,4 presents the summary of revised values for the neutrino
counting rates in the Cl- and Ga-experiments, along with the flux
values from SK and SNO \cite{Fukuda}. These data allow to conclude that
the observed energy spectrum of the electron neutrinos, normalized
to SSM, demonstrates a decrease of the neutrino flux relative to
SSM with an increase of their energy. This decrease occurs
nonlinearly, since the most rapid decay is observed in the
intermediate energy range.

\begin{table}[htbp]
\begin{tabular}{|p{95pt}|p{65pt}|p{65pt}|p{65pt}|}
\hline
Experiment& 
Data \par (SNU)& 
SSM \par flux \par (SNU)& 
Data (in  \par SSM units) \\
\hline
SAGE ($^{71}$Ga) I&
67.1 $\pm $ 7.9& 
129 $\pm $ 8 & 
0.520 $\pm $ 0.061 \\
~~~~~~~~~~~~~~~~~~~II&
66.1 $\pm $ 5.3& - &0.512 $\pm $ 0.041 \\
\hline
GNO ($^{71}$Ga)& 
62.9 $\pm $ 6.0& 
129 $\pm $ 8& 
0.488 $\pm $ 0.042 \\
\hline
Homestake ($^{37}$Cl)& 
3.03 $\pm $ 0.22& 
7.7 $\pm $1.2& 
0.393 $\pm $ 0.029 \\
\hline
Super Kamiokande (H$_{2}$O)& 
2.39$\pm $0.07 \par (10$^{6}$cm$^{-2}$s$^{-1})$& 
5.15$\pm $0.72 \par (10$^{6}$cm$^{-2}$s$^{-1})$& 
0.465 $\pm $ 0.014 \\
\hline
SNO (D$_{2}$O)& 
1.80$\pm $0.11 \par (10$^{6}$cm$^{-2}$s$^{-1})$& 
-& 
0.349 $\pm $ 0.022 \\
\hline
\end{tabular}
\caption{
Revised values for the neutrino flux
from the GALLEX, GNO, SAGE and 
Super Camiokande (SK) experiments. 
}
\end{table}

It seems worthwhile to evaluate the flux of pp-neutrinos,
constituting a main part of the solar neutrino spectrum, using
the experimental data listed in Table\,4.

Let us adopt the neutrino counting rate in the Ga-experiment to be
$66.1\pm 5.3$\,SNU, as having the least error value. This counting
rate is composed of all the components of the solar neutrino flux:
pp+$^8B+F_{\rm o}(Ga)$, where $F_{\rm o}(Ga)={}^7Be+GNO+pep+hep$, and the
subscript ``o'' shows that this is an observed value. Evaluation
of errors will be discussed later.

Using the SNO flux and the cross section for $^8$B neutrinos in
the SSM form, $(2.4\pm 0.6)\times 10^{-42}$\,cm$^{-2}$, we find
the contribution of $^8$B neutrinos to the Ga-experiment to be
4.3\,SNU. Subtracting this value from the total counting rate of
the Ga-experiment, one gets $pp+F_{\rm o}(Ga)=61.8$\,SNU.

The observed counting rate in the Cl-experiment is $^8B+F_{\rm
o}(Cl)=3.03\pm 0.22$\,SNU. Using again the SNO flux and the
cross section in SSM, $(1.14\pm 0.04)\times 10^{-42}$\,cm$^{-2}$,
we find the contribution of $^8$B neutrinos to the Cl-experiment
to be 2.05\,SNU. Thus, F$_{\rm o}$(Cl)=0.98\,SNU. Taking after
Abdurashitov et al. (2002) that the energy-dependent surviving
factor with respect to SSM, F$_{\rm o}$(Ga)/F$_{\rm t}$(Ga), is
the same as in the Cl-experiment, F$_{\rm o}$(Cl)/F$_{\rm
t}$(Cl)=0.527, we get F$_{\rm o}$(Ga)=0.527F$_{\rm t}$(Ga)=24.4\,SNU.
However, a transition surviving factor in this
form essentially depends on the flux ratio of the SNO- to
Cl-neutrinos in the SSM. To diminish this influence, we multiplied
the transition factor by the value of this ratio. Then, F$_{\rm
o}$(Ga)=0.47 F$_{\rm t}$(Ga)=21.7\,SNU. Thus, the observed
pp-counting rate in the Ga-experiment is 40.1\,SNU. The results
obtained above are collected in Table\,5.

\begin{table}[htbp]
\begin{tabular}{|c|c|c|}
\hline
Neutrino source & 
Cl (SNU)& 
Ga (SNU) \\
\hline
PP& 
0& 
40.1$\pm $15.6 \\
\hline
$^{8}$B& 
2.05$\pm $0.20 & 
4.3$\pm $1.0 \\
\hline
F$^{\ast }$& 
0.98$\pm $0.23 & 
21.7$\pm $8.7  \\
\hline
sum& 
3.03$\pm $0.22 & 
66.1$\pm $5.3  \\
\hline
\multicolumn{3}{|c|}{F$^{\ast }$=$^{7}$Be + CNO + pep + hep}  \\
\hline
\end{tabular}
\caption{
Values for the neutrino counting rate 
in the Cl- and Ga- experiments. 
}
\end{table}
 
Similarly, we have estimated the pp-counting rate 
for the maximal flux values  in the
Cl- and Ga-experiments as 42.1\,SNU.

In order to get the observed pp-flux of electron neutrinos on the
Earth, one can divide the obtained counting rate by the SSM cross
section for electron neutrinos, $(11.7\pm 0.3)\times
10^{-46}$\,cm$^2$, thus getting $(3.4\pm1.3)\times
10^{10}$\,cm$^2$c$^{-1}$ and $(3.6\pm1.3)\times
10^{10}$\,cm$^2$c$^{-1}$, respectively.

For more precise evaluation of the observed pp-flux it is
necessary to measure $^7$Be flux, which constitutes a major part
of F$_{\rm o}$(Ga), and, naturally, to diminish the influence of
the statistical and systematical errors in the Ga-experiments.

\section{Conclusions}

The analysis of power spectra for the series of neutrino
experiments has revealed that they display, 
11-, 5-, 2- and 1.3-year quasi-periodicities, which are also
observed in the solar activity. The most stable one turns out to
be a quasi-5-year fluctuation of the neutrino flux detected in the
radiochemical experiments. Quantitative estimates of the amplitude
of these fluctuations have shown that the variable component of
the neutrino flux deficit is likely to be connected with cyclic
changes of the solar magnetic field.

We have considered two alternative possibilities to explain an
origin of the variable component of the solar neutrino deficit.
The first possibility implies real changes of the solar neutrino
flux, while the second one suggests that the observed fluctuations
are due to effect of an unknown factor, which operates
synchronously with the solar activity and  influences the results
of neutrino experiments. The second possibility is supported by
the fact that the solar activity seems to produce an opposite
effect on the detection of neutrino flux in the GAL+GNO and SAGE
experiments (Table\,3). In order to check this assertion, it might
be useful, in particular, to perform synchronous observations in
SAGE and GNO. In any case, the maximum deviation from the averaged
neutrino flux is $\pm 0.1$\,SSM.

We argue that earlier reported average neutrino flux values
derived with the radiochemical experiments need corrections. The
revised values, given in Table\,4, turn out to be larger for the
Homestake data, and smaller for the SAGE data, than those
previously published. The revised counting rates have been used to
estimate the observed pp-flux of the electron solar neutrino at
the Earth as $0.57\pm 0.22$\,SSM.

Any further refinement of the values presented in Tables\,4 and 5
as well as a study of variable component of the solar neutrino
flux deficit, require, in particular, significant reduction of the
systematic and statistical errors in the neutrino experiments (see
Tables\,2, 3).

\acknowledgements

The authors are grateful to Dr. N. Beskrovnaya for her 
assistance in preparation of this manuscript.
The work was supported by the Program of the Presidium of the
Russian Academy of Sciences ``Non-stationary phenomena in
astronomy''.

\end{article}
\end{document}